# Influence of Mutual Drag of Light and Heavy Holes on Magnetoresistivity and Hall-effect of *p*-Silicon and *p*-Germanium

**I.I. Boiko[1]**

*Institute of Semiconductor Physics, NAS of Ukraine, 45, prospect Nauky, 03028 Kyiv, Ukraine*

Hall-effect and magnetoresistivity of holes in silicon and germanium are considered with due regard for mutual drag of light and heavy band carriers. Search of contribution of this drag shows that this interaction has a sufficient influence on both effects.

## 1. Introduction

In previous work (see Ref. [1]) we have investigated the influence of mutual drag of heavy and light holes on conductivity of *p*-germanium and *p*-silicon. It was shown that this drag significantly diminishes total conductivity of crystal. On practice this effect escaped attention of great number of earlier investigators. One of the reason was vain attempt to describe the mutual drag of band carriers belonging to different groups (see, for example, Refs. [2, 3]) with the help of widespread tau-approximation in the course of solution of kinetic equation (see Refs. [4, 5]). In this paper we use the method of balance equation (see Refs. [6, 7, 8]), which reliably allows to introduce in consideration the mutual drag of carriers from two bands which close up in the center of wave vector space. For simplicity of calculations we accept here spherical bands approximation. So dispersion laws of holes have the simple form:

$$\varepsilon_{\vec{k}}^{(a)} = \hbar^2 k^2 / 2m_a \ . \quad (a = 1 \text{ or } 2) \tag{1}$$

In this formula $m_a$ are effective masses ($m_1$ is mass of light holes and $m_2$ is mass of heavy holes). The approximation (1) is acceptable for considered here situations.

## 2. Balance equations

Consider the set of two balance equations obtained as a first momentum of quantum kinetic equations (see Refs. [7, 9]):

[1] E-mail: igorboiko@yandex.ru

$$e[\vec{E} + (1/c)(\vec{H} \times \vec{u}^{(a)})] + \vec{F}^{(a)} + \sum_{b=1}^{2} \vec{F}^{(a,b)} = 0 \,. \qquad (a = 1, 2) \qquad (2)$$

Here vectors $\vec{E}$ and $\vec{H}$ represent electrical and magnetic fields, the values $\vec{u}^{(a)}$ are drift velocities of light and heavy holes, $\vec{F}^{(a)}$ is resistant force related to external scattering system, $\vec{F}^{(a,b)}$ is force connected with Coulomb interaction of heavy and light holes.

We restrict here our consideration by external scattering system containing charged impurities and acoustic phonons.

Accepting the model of non-equilibrium distribution functions of holes as Fermi functions with argument containing for $a$-group the shift of velocity $\vec{v}^{(a)}(\vec{k}) = \hbar^{-1}(\partial \varepsilon_{\vec{k}}^{(a)} / \partial \vec{k})$ on correspondent velocity $\vec{u}^{(a)}$

$$f_{\vec{k}}^{(a)} = f^{0(a)}(\vec{v}^{(a)}(\vec{k}) - \vec{u}^{(a)}), \quad (a = 1, 2) \qquad (3)$$

(here $f^{0(a)}(\vec{v}^{(a)}(\vec{k}))$ is equilibrium distribution function for $a$-carriers) we obtain the following expressions for forces presented in Eq. (2):

$$\vec{F}^{(a)} = -e\,\beta^{(a)}\vec{u}^{(a)} \;\; ; \quad \vec{F}^{(a,b)} = -e\,\xi^{(a,b)}\left(\vec{u}^{(a)} - \vec{u}^{(b)}\right). \qquad (4)$$

Farther we consider only undegenerate band carriers. Then (see Ref. [10])

$$\beta^{(a)} = \lambda^{(a)} + \chi^{(a)} \; ; \qquad (5)$$

$$\lambda^{(a)} = \frac{4\sqrt{2\pi m_a}\,e^3\,n_I}{3(k_B T)^{3/2}\varepsilon_L^2} \int_0^\infty \frac{q^3 dq}{(q^2 + q_0^2)^2} \exp\left(-\frac{\hbar^2 q^2}{8 m_a k_B T}\right) \; ; \qquad (6)$$

$$\chi^{(a)} = \frac{8\sqrt{2}\,\Xi_A^2 (k_B T)^{3/2} m_a^{5/2}}{3\pi^{3/2}\hbar^4 e\,\rho\,s^2} \; ; \qquad (7)$$

$$\xi^{(a,b)} = \frac{8\gamma\,e^3\,m_b^2\,p_a}{3 k_B T\,\hbar\,m_a} \int_0^\infty \frac{q^2 dq}{(q^2 + q_0^2)^2} \exp\left[-\frac{\hbar^2 q^2}{8 k_B T}\left(\frac{1}{m_1} + \frac{1}{m_2}\right)\right] \; ; \qquad (8)$$

$$\gamma = \int_{-\infty}^{\infty} \frac{w^2 dw}{\sinh^2 w} \approx 3.29 \; ; \qquad (9)$$

$$q_0^2 = \frac{4\pi\,e^2(p_1 + p_2)}{\varepsilon_L\,k_B T} \quad . \qquad (10)$$

Here $p_a$ is density of $a$-holes, $n_I$ is density of charged impurities, $\Xi_A$ is deformation potential.

We assume $n_I = p = p_1 + p_2$.

For nondegenerate carriers

$$p_1 / p_2 = (m_1 / m_2)^{3/2} = w . \qquad (11)$$

For $p$-germanium $w = 0.042$, for $p$-silicon $w = 0.153$ (see Ref. [5]).

From the formulae (2) and (4) one obtains the system of equations for drift velocities:

$$\vec{E} + (1/c)(\vec{H} \times \vec{u}^{(a)}) - \beta^{(a)} \vec{u}^{(a)} - \sum_{b=1}^{2} \xi^{(a,b)}\left(\vec{u}^{(a)} - \vec{u}^{(b)}\right) = 0 . \qquad (a = 1, 2) \quad (12)$$

The total density of current is

$$\vec{j} = \sum_{a=1}^{2} \vec{j}^{(a)} = e\sum_{a=1}^{2} p_a \vec{u}^{(a)} . \qquad (13)$$

## 3. **Hall-coefficient and Magnetoresistivity**

Consider that case when external magnetic field $\vec{H}$ and total current $\vec{j}$ are directed along $z$-axis and $x$-axis consequently:

$$\vec{H} = (0, 0, H_z) ; \ \vec{j} = (j_x, 0, 0) . \qquad (14)$$

Then electrical field $\vec{E}$ has the following components:

$$\vec{E} = (E_x, E_y, 0) . \qquad (15)$$

Here the component $E_x$ is related to applied electrical field; $E_y$ is Hall-component.

Represent the relation between measured components of total current and electrical field in the form

$$j_x = \sigma * (H)E_x . \qquad (16)$$

The scalar value $\sigma * (H)$ we call as magnetoresistivity.

Introduce Hall-coefficient $R_H$ by the following relation:

$$R_H = \left| \frac{E_y}{H_z j_x} \right| . \qquad (17)$$

To calculate the values $\sigma * (H)$ and $R_H$ we should to solve the system of equations (12) on the first stage. For the case (14), (15) we have the system of five equations:

$$\vec{E} + (1/c)(\vec{H} \times \vec{u}^{(1)}) = \beta^{(1)} \vec{u}^{(1)} + \xi\left(\vec{u}^{(1)} - \vec{u}^{(2)}\right) ;$$

$$\vec{E} + (1/c)\,(\vec{H} \times \vec{u}^{(2)}) = \beta^{(2)}\,\vec{u}^{(2)} + \xi\,w^{-1}\!\left(\vec{u}^{(2)} - \vec{u}^{(1)}\right) \quad ; \tag{18}$$

$$p_1 u_y^{(1)} + p_2 u_y^{(2)} = 0$$

Here $\xi = \xi^{(1,2)}$ (see Eqs. (8) and (11)); $\vec{u}^{(a)} = (u_x^{(a)}, u_y^{(a)}, 0)$ ; $(a = 1, 2)$. The unknown values are $u_x^{(1)}$, $u_y^{(1)}$, $u_x^{(2)}$, $u_y^{(2)}$ and $E_y$.

Analytical solution of the system (18) is very simple but rather complicated in the form. Therefore we present here results of our numerical calculations only by figures (here $H_0 = c\beta^{(1)}$). To carry the calculations we used the following numerical values:

$$\varepsilon_L = 12 \;,\; \rho\,s^2 = 1.66\cdot10^{11}\,Pa \;,\; \Xi_A = -4.2\,eV \;\; \text{for } p\text{-silicon and}$$

$$\varepsilon_L = 16 \;,\; \rho\,s^2 = 1.26\cdot10^{11}\,Pa \;,\; \Xi_A = 1.9\,eV \;\; \text{for } p\text{-germanium.}$$

## 4. Results of Calculations

Fig. 1(a, b) gives the possibility to compare Hall-coefficients calculated for the case when interband drag is taken into account (see $R_H(H, \xi)$) and is not taken (see $R_H(H, 0)$). One can notice that $R_H(H, \xi) < R_H(H, 0)$ and the difference is especially significant at small magnetic field.

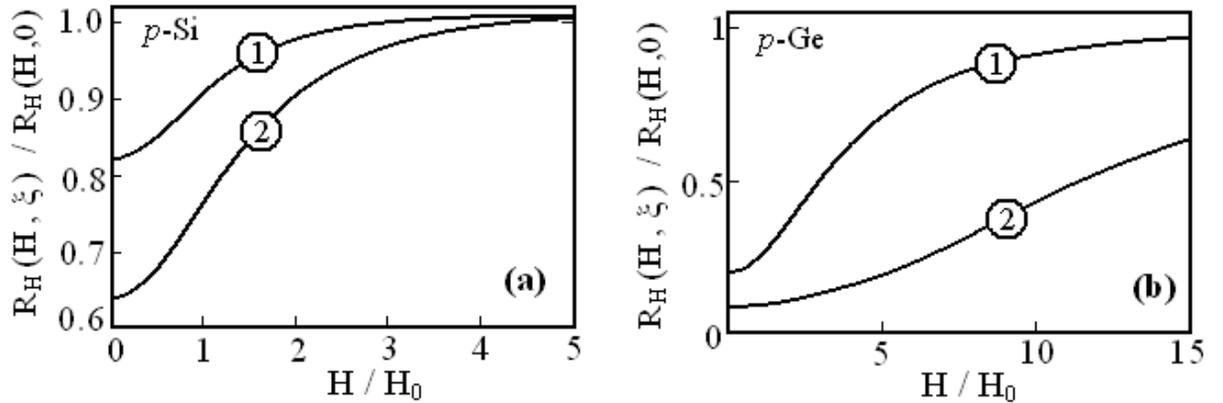

Fig. 1(a, b). Influence of interband drag on Hall-coefficient.

(a) $p$-silicon: $p = 10^{17}$ cm$^{-3}$ ; (b) $p$-germanium: $p = 10^{14}$ cm$^{-3}$ ;

$1 - T = 50$ K , $2 - T = 100$ K .

Figs. 2(a, b) and 3 show dependence of Hall-coefficient on intensity of magnetic field. It follows that drag makes this dependence more smooth (in some cases this dependence practically disappear; see Fig. 3, the curve 2).

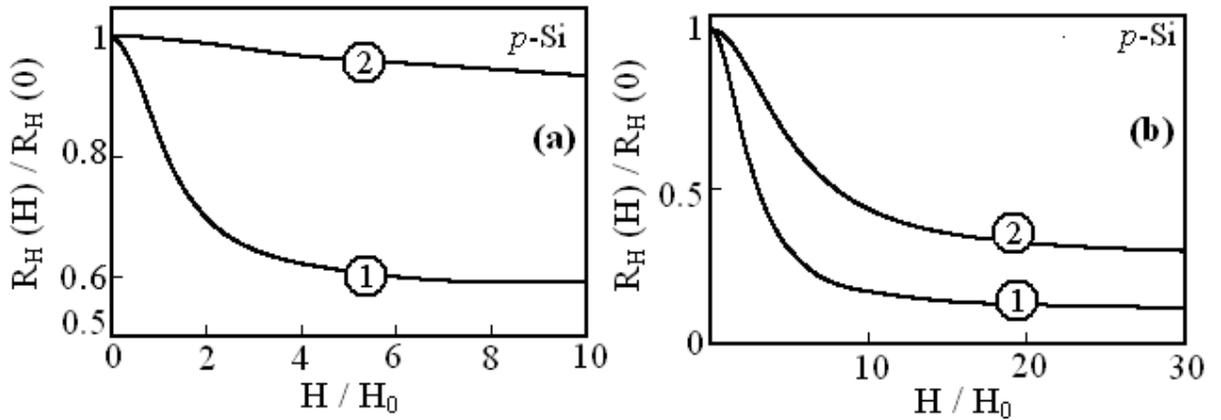

Fig. 2 (a, b). Dependence of relative Hall-coefficient of *p*-silicon on dimensionless magnetic field. $p = 10^{14}$ cm$^{-3}$. (a): T = 50 K ; (b): T = 100 K .

$1 - \xi = 0$ , $2 - \xi \neq 0$ .

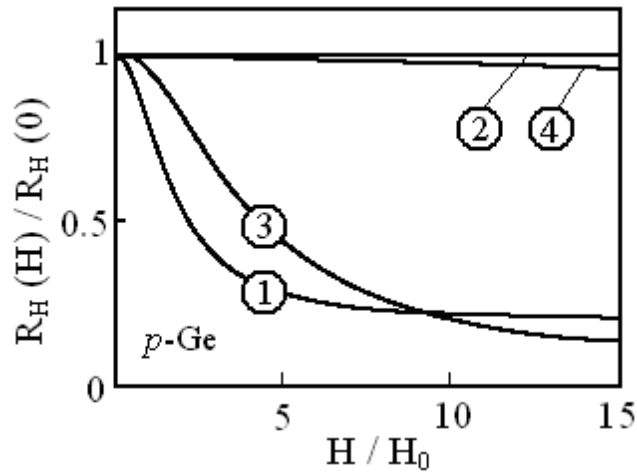

Fig. 3. Dependence of relative Hall-coefficient of *p*-germanium on dimensionless magnetic field. $p = 10^{14}$ cm$^{-3}$ .

T = 50 K . $1 - \xi = 0$ , $2 - \xi \neq 0$ .

T = 100 K : $3 - \xi = 0$ , $4 - \xi \neq 0$ .

Figs. 4 (a, b) and 5 (a, b) represent the dependence of relative magnetoconductivity of *p*-silicon and *p*-germanium on dimensionless magnetic field. Calculation show that influence of intensity of the field on conductivity is significantly moderated by interband drag (especially in germanium where difference between effective masses of light and heavy holes is apartly high).

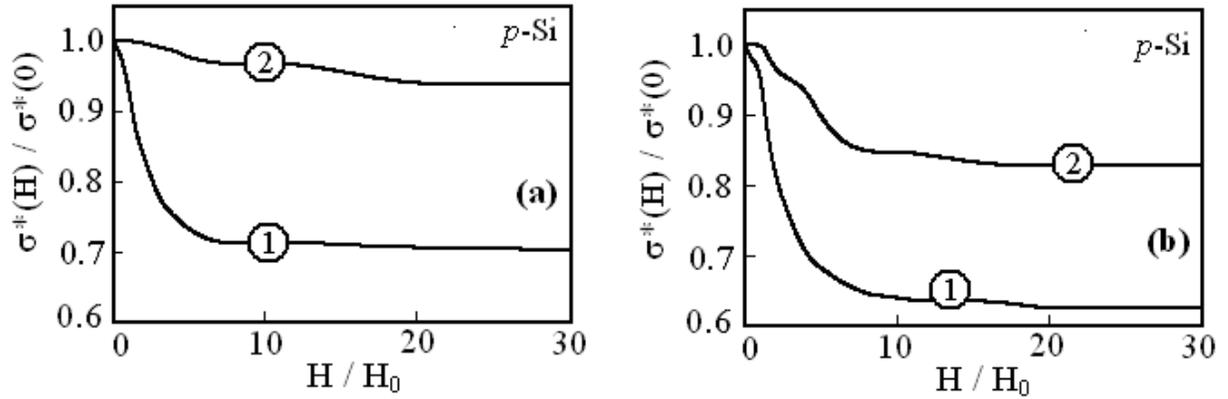

Fig. 4 (a, b).

Dependence of relative magnetoconductivity of *p*-Si on dimensionless magnetic field.

$p = 10^{14}\,cm^{-3}$ . (a) : $T = 50\,K$ ; (b): $T = 100\,K$ . $1 - \xi = 0$ , $2 - \xi \neq 0$ .

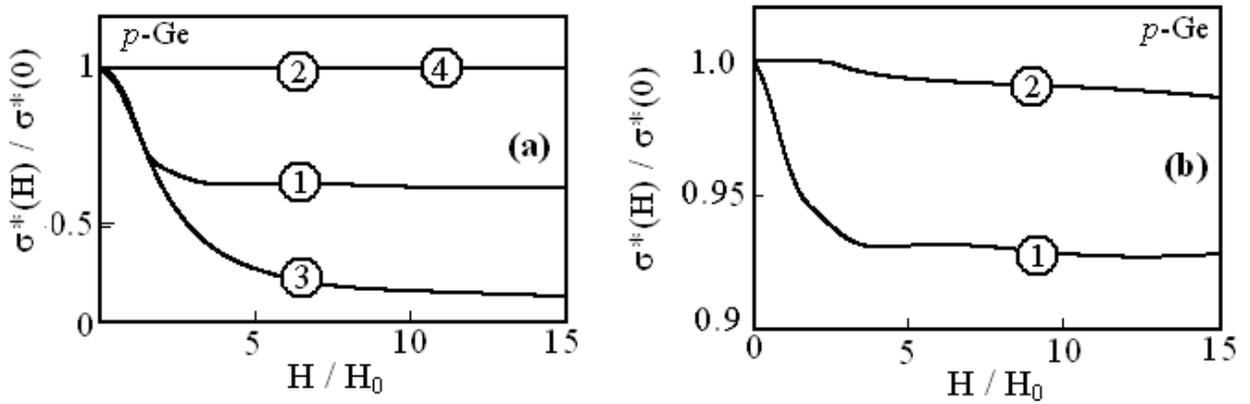

Fig. 5 (a, b).

Dependence of relative magnetoconductivity of *p*-Ge on dimensionless magnetic field.

(a): $p = 10^{14}\,cm^{-3}$ . $T = 50\,K$ : $1 - \xi = 0$ , $2 - \xi \neq 0$ . $T = 100\,K$ : $3 - \xi = 0$ , $4 - \xi \neq 0$ .

(b): $p = 10^{17}\,cm^{-3}$ ; $T = 100\,K$ . : $1 - \xi = 0$ , $2 - \xi \neq 0$ .

---